# High-brightness seven-octave carrier envelope phase-stable light source

Ugaitz Elu[1], Luke Maidment[1], Lenard Vamos[1], Francesco Tani[2], David Novoa[2], Michael H. Frosz[2], Valeriy Badikov[3], Dmitrii Badikov[3], Valentin Petrov[4], Philip St. J. Russell[2,5], Jens Biegert[1,6,*]

[1]ICFO - Institut de Ciencies Fotoniques, The Barcelona Institute of Science and Technology, 08860 Castelldefels, Barcelona, Spain.
[2]Max-Planck Institute for Science of Light, Staudtstraße 2, 91058 Erlangen, Germany
[3]High Technologies Laboratory, Kuban State University, Stavropolskaya Str. 149, 350040 Krasnodar, Russia
[4]Max-Born-Institute for Nonlinear Optics and Ultrafast Spectroscopy, 2A Max-Born-Str., D-12489 Berlin, Germany
[5]Department of Physics, Friedrich-Alexander-Universität, Staudtstraße 2, 91058 Erlangen, Germany
[6]ICREA, Pg. Lluis Companys 23, 08010 Barcelona, Spain.
e-mail: jens.biegert@icfo.eu

**High-brightness sources of coherent and few-cycle-duration light waveforms with spectral coverage from the UV to the THz would offer unprecedented versatility and opportunities for a spectacular range of applications from bio-chemical sensing[1], to time-resolved and nonlinear spectroscopy, to attosecond light-wave electronics[2,3]. Combinations of various sources with frequency conversion[4,5] and supercontinuum generation[6–9] can provide relatively large spectral coverage, but many applications require much broader spectral range[10] and low-jitter synchronization for time-domain measurements[11]. Here, we present a carrier-envelope-phase stable light source, seeded by a mid-IR frequency comb[12,13], with simultaneous spectral coverage across 7 optical octaves, from the UV (340 nm) into the THz (40,000 nm). Combining soliton self-compression and dispersive wave generation in an anti-resonant-reflection photonic crystal fibre with intra-pulse difference frequency generation in $BaGa_2GeSe_6$, the spectral brightness is 2-5 orders of magnitude above synchrotron sources. This enables high-dynamic-range spectroscopies and provides enticing prospects for attosecond physics and material sciences[14,15].**

Spectrally broad and coherent light sources are indispensable tools that enable ultrasensitive spectroscopy and imaging of gases, plasmas, liquids and solids[16]. For instance, a broad spectral coverage provides access to multiple absorption signatures without the need for numerous individual coherent light sources[17], and without mechanical tuning. Hyperspectral spectroscopy and imaging[18] with visible to THz radiation the rapid and secure discrimination of objects in applications as diverse as cultural heritage[19], security or food industry, the identification of greenhouses gases, pollutants and harmful substances, or the identification of cancer markers and DNA[20]. The spectacular wealth of information available from molecular spectroscopy extends from the composition of a molecule, to its rotational and vibrational spectra, and to the coherent and incoherent couplings between various electronic states. Phase-stable ultrashort pulses in the mid-IR to THz permit the interrogation of valley population in 2D materials[21], or creation of light-induced exotic states of matter such as topological phases[22] and superconductors[15]. Clearly, there exist a vast range of spectacular possibilities (see Fig. 1a); however, many of them place high demands on the light source, requiring large spectral coverage, high brilliance and spectral brightness, short pulse duration, carrier-envelope-phase stability, and any combination thereof.

The generation of coherent broadband radiation combining such features remains a challenging task due to i) the lack of efficient laser gain media in the infrared, ii) the limitations of quadratic nonlinear media to phase-match and transmit across a vast spectral range, and iii) the scaling of third order processes to long wavelengths. On the short wavelength side, self-phase modulation (SPM) and soliton dynamics have been exploited in numerous ways as efficient means to generate UV radiation[23,24], and to yield short optical pulses, with several percent efficiency. Depending on the available pulse energy and peak power, implementations employ filamentation in gases and solids, gas-filled hollow-core fibres, photonic crystal fibres, or other highly nonlinear media. Ultrabroadband radiation in the mid-IR to THz is readily generated by difference frequency generation (DFG) with a quadratic nonlinearity, thus providing much higher efficiency compared with a third order process like SPM. Inter-pulse DFG[12] of two pulses with different centre frequencies, but with a common origin, or intra-pulse DFG (IP-DFG) of a broadband intense pulse[25], are reliable implementations for generating mid-IR to THz radiation. Both methods provide passively carrier-envelope-phase (CEP) stable frequency combs[26,27], which are used as the frontend for mid-IR OPCPA[12,13], or enable investigations that depend on the stability of the pulse's electric field waveform, e.g. in attosecond physics, for high harmonic generation (HHG), electron diffraction or electro-optical sampling (EOS).

Here, we build on our recent work[13] in which we have shown that a gas-filled anti-resonant-reflection photonic crystal fibre (ARR-PCF) achieves highly efficient soliton self-compression of a 10-W CEP-stable mid-IR pulse to the single-cycle limit. We show that pressure control and ARR-PCF design concomitantly accomplish efficient spectral broadening and energy transfer into a dispersive wave (DW) in the UV. The high spectral brightness of this process, together with the soliton-compressed mid-IR part of the spectrum, are leveraged as high peak-power seed components for IP-DFG in $BaGa_2GeSe_6$ (BGGSe)[28], which is a highly-efficient new nonlinear-medium for mid-IR to THz frequency conversion[29]. Figure 1b shows the simple experimental layout in which we combine the ARR-PCF with IP-DFG, resulting in a simultaneous 7-octave-spanning (340 nm to 40000 nm) coherent output with record spectral brightness and without switching between different nonlinear-crystals[30]. The resulting continuum is CEP stable and under 3-optical cycles in duration.

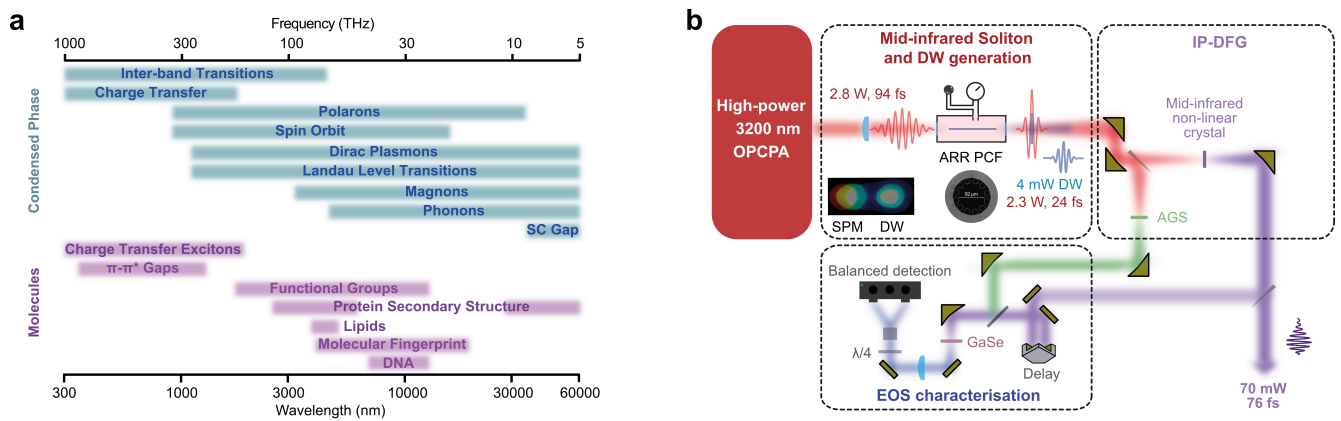

Fig. 1. Measurement application and layout of the seven-octave coherent light source. (a) Possible applications in quantum light science for condensed phase and molecular phase investigations are shown. (b) Schematic of the system layout. A high-power mid-IR OPCPA seeds the 92-µm core diameter and 20-cm long gas-filled ARR-PCF with CEP-stable pulses at 160 kHz. Sub-cycle soliton-self compression and DW generation yield a CEP-stable supercontinuum. High-brightness IP-DFG extends the wavelength range of the comb into the THz spectral range, up to 40,000 nm. Also indicated is the EOS metrology for direct measurement of the electric-field waveform and corresponding spectrum.

Pulses with 94-fs duration and centre wavelength of 3200 nm at 160 kHz repetition rate and 2.8 W average power from a passively CEP-stable mid-IR OPCPA[13] are coupled into a 20-cm long ARR-PCF fibre with core diameter of 92 µm (93% launching efficiency). The fibre is mounted inside a gas cell equipped with 3-mm-thick $CaF_2$ windows and filled with Ar at pressures up to 35 bars. Varying the gas pressure changes its dispersion, thus influencing the self-phase modulation (SPM) process and pulse propagation. Figure 2 shows the temporal and spectral evolution of the emerging pulse, at different pressures, measured with second-harmonic frequency-resolved optical gating (SHG-FROG). Clearly visible in Fig. 2a is the onset of temporal compression above 20 bar, which occurs when the anomalous dispersion of the gas increases sufficiently to balance the nonlinear dispersion from SPM[24]. Soliton-self compression at 25 bar yields a near-single-cycle temporal waveform in the mid-IR. Figure 2b shows the associated spectrum together with results from a carrier-resolved unidirectional nonlinear wave-equation model[31]. Based on the excellent match between theory and experiment, we can confidently predict the temporal structure (Fig. 2c) corresponding to the entire continuum, from 340 to 5000 nm, in Fig. 2b. Further increase in pressure to 30 bar results in the generation of a temporal shock front with sub-cycle duration (Fig. 2c), concomitant with the emergence of the DW[24] (Fig. 2b). The measured mid-IR pulse (Fig. 2a) exhibits multiple temporal peaks with widely varying temporal phase. At 35 bar, we generated an average power of 5 mW (31 nJ) in the DW, corresponding to an energy transfer of 0.2%. Its signature in the temporal domain (Fig. 2c) are high frequency modulations at the trailing edge of the pulse and the corresponding supercontinuum is broadest (Fig. 2b), ranging from 340 nm to 5000 nm.

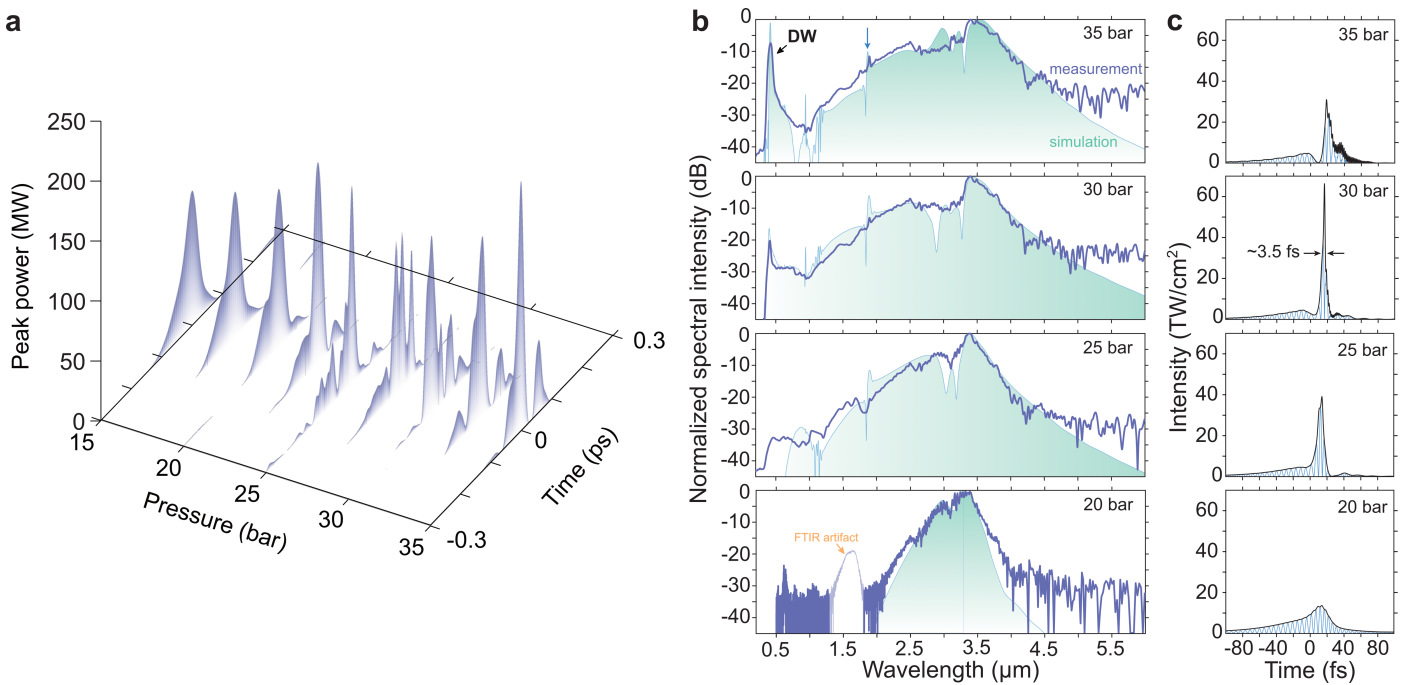

Fig. 2. Pressure dependence of the mid-IR soliton self-compression and DW generation. (a) Measured temporal intensity profiles from SHG FROG. (b) Measured spectra (dark blue) together with simulation results (petrol). The petrol-coloured arrow indicates a fibre resonance (c) Predicted temporal structures for the simulated spectra in (b). At 25 bar the pulses are compressed below 2 optical cycles with peak power exceeding 200 MW. Increasing pressure to 35 bar, the DW is most pronounced and the pulse reaches a peak power close to 250 MW with the main temporal structure compressed to below 2 optical cycles.

The evolution of spectra and temporal profiles (Fig. 2) provides key insight into the optimisation of the soliton dynamics in the ARR-PCF for broadband and efficient IP-DFG. Our pressure scan reveals that a pressure of 25 bar provides the best balance between a clean pulse structure and "narrow-enough" spectrum with high spectral intensity for efficient IP-DFG. Thus, a pressure range between 20 and 35 bar is expected to provide both broad supercontinuum generation to the UV range, and efficient IP-DFG into the THz range. Before corroborating this, we examine the suitability of three mid-IR nonlinear crystals for IP-DFG: a 1-mm-thick uncoated GaSe, 2-mm-thick AR-coated ZGP and 2.6-mm-thick AR-coated BGGSe. After the ARR-PCF, the emerging pulse is focused with a 50-mm off axis paraboloid to a beam waist of 200 µm into the IP-DFG crystal and the generated waveform is measured with EOS. Figure 3a shows that the spectrum generated from ZGP is centered at shorter wavelengths compared with GaSe. The best power distribution for broadest spectral coverage is achieved with BGGSe, despite the thicker crystal. All crystals yield electric-field waveforms with durations from 2 to 2.5 optical cycles (see Fig. 3b) which are CEP stable. The ZGP crystal generates 70 mW (0.45 µJ) of power in a 76-fs-duration (2.3 optical cycles) pulse whose spectrum is centred at 9.9 µm. This is nearly two orders of magnitude higher power (see Fig. 3c) than from GaSe, which yields 1.8 mW (11.25 nJ) of power in a 96-fs-duration (2.2 optical cycles) pulse whose spectrum is centred at 13.1 µm. BGGSe clearly outperforms both crystals in its combined figure of merit: We measure a 86-fs-duration pulse (2.5 optical cycles) with 25 mW (0.16 µJ) of power, centred at 10.3 µm. We note that 2-3 times higher power could be achieved with custom coated optics.

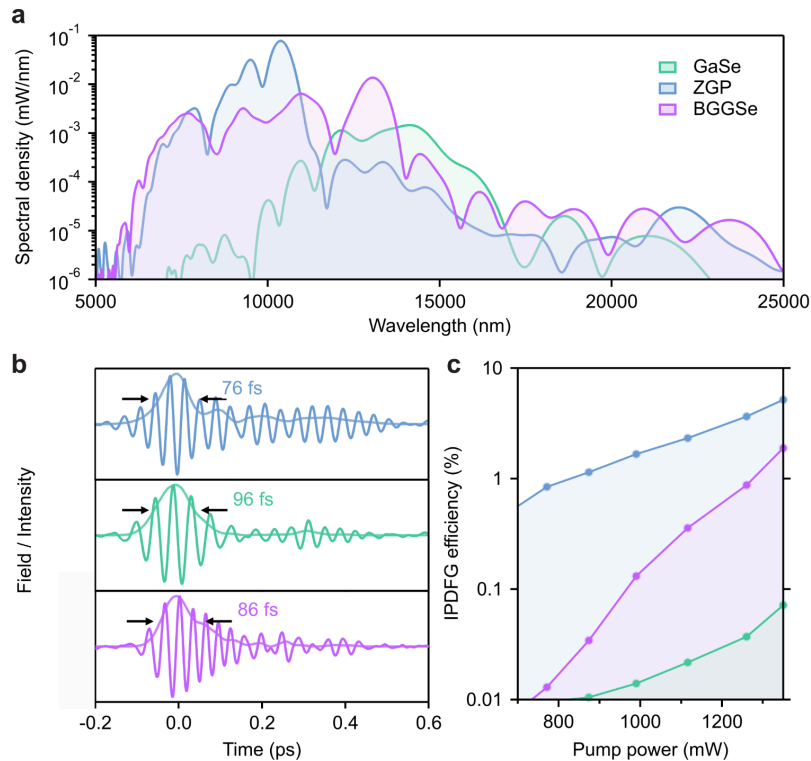

Fig. 3. IP-DFG comparison at 25 bar between ZGP, BGGSe and GaSe. (a) Measured spectral density generated in ZGP (blue), BGGSe (violet) and GaSe (petrol). (b) Electric field waveforms and intensity profiles measured with EOS for the three crystals in (a). The shortest pulse duration is found for ZGP (blue) due to narrower spectrum, which is less sensitive to residual uncompensated dispersion. The "cleanest" electric field waveform is measured from GaSe. We attribute this to the red-shifted spectrum and the correspondingly lower sensitivity to residual dispersion. (c) shows how the IP-DFG efficiency scales for the three crystals. BGGSe provides an excellent balance of efficiency and achievable temporal waveform.

The generation of a few-cycle pulse at an IP-DFG efficiency of up to 2% makes BGGSe an ideal choice for the generation of coherent and CEP-stable mid-IR supercontinua. Adjustment of the ARR-PCF pressure to 35 bar results in peak powers as high as 2.5 MW with a spectral brightness in the UV (between 340 and 450 nm) comparable to the mid-IR pump. A lower pressure of 20 bar provides simultaneous generation of wavelengths up to 40,000 nm with THz peak power of 1.8 MW. Our table top-source constitutes a coherent and ultrabroadband comb covering 7 optical octaves from 340 nm to 40,000 nm, with electric field waveforms that are CEP stable and that correspond to sub-3-optical cycle pulses.

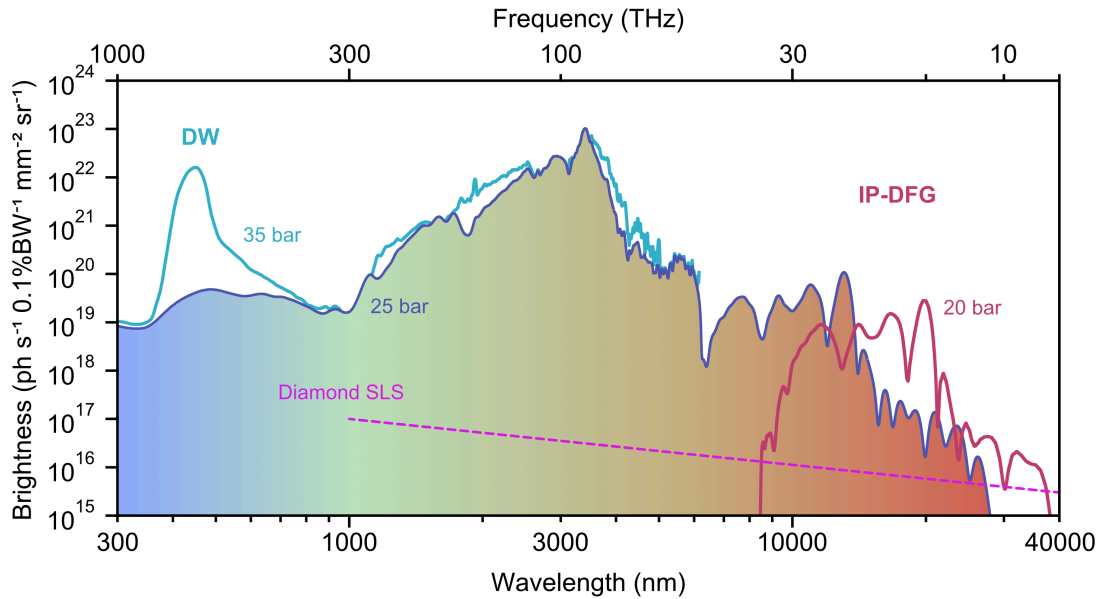

Fig. 4: High-brightness 7-octave few-cycle supercontinuum. The gradient plot shows the generated supercontinuum for an ARR-PCF pressure of 25 bar. For comparison, values from the Diamond Synchrotron Light Source are overlaid, clearly showing the much higher infrared performance of our table-top source. In addition, we show how pressure tuning can optimize the UV and THz spectral density. At 35 bar of pressure, the DW reaches values only 15 dBm/nm below the pump wavelength of 3200 nm. A low pressure of 20 bar permits optimizing for the generation of wavelength up to 40,000 nm.

In summary, we have presented a high-brightness table-top source of coherent CEP-stable waveforms whose spectrum extends from 340 nm to 40,000 nm, without the need to change nonlinear crystals. Pressure-tuning permits adjustment of the spectral intensity, depending on experimental need. UV peak powers up to 2.5 MW and THz peak powers of 1.8 MW open up entirely new prospects for nonlinear and multi-dimensional spectroscopies in the time domain due to the stability and sub-3-cycle duration of the electric field waveforms. The seven-octave-wide coherent spectrum meets the need for tunable sources without spectral gaps, and it empowers new multi-modal measurement methodologies, for instance in molecular spectroscopy, physical chemistry or solid-state physics. The high peak power permits peak intensities in the TW/cm$^2$ range in the UV and in the THz spectral range to be reached, thus enabling strong field and attosecond science and allowing exploration of new phenomena such as light-induced phase transitions, superconductivity or topological physics.

**Acknowledgement**
J.B. and group acknowledge financial support from the European Research Council for ERC Advanced Grant "TRANSFORMER" (788218) and ERC Proof of Concept Grant "miniX" (840010), the European Union's Horizon 2020 for FET-OPEN "PETACom" (829153), FET-OPEN "OPTOlogic" (899794), Laserlab-Europe (EU-H2020 654148), Marie Sklodowska-Curie grant No 860553 ("Smart-X"), MINECO for Plan Nacional FIS2017-89536-P; AGAUR for 2017 SGR 1639, MINECO for "Severo Ochoa" (SEV- 2015-0522), Fundació Cellex Barcelona, CERCA Programme / Generalitat de Catalunya, the Alexander von Humboldt Foundation for the Friedrich Wilhelm Bessel Prize. We thank I. Tyulnev and M. Enders for their assistance.